\documentstyle[12pt]{article}
\title{Long-run behavior of games with many players} 
\author{Jacek Mi\c{e}kisz \\ Institute of Applied Mathematics \\
and Mechanics \\ Warsaw University  \\ ul. Banacha 2  \\ 02-097
Warsaw, Poland 
\\ e-mail: miekisz@mimuw.edu.pl} 
\begin{document} 
\baselineskip=20pt
\maketitle 

\noindent {\bf Abstract}: We discuss similarities and differencies between systems 
of many interacting players maximizing their individual payoffs and particles minimizing 
their interaction energy. We analyze long-run behavior of stochastic dynamics
of many interacting agents in spatial and adaptive population games. 
We review results concerning the effect of the number of players and the noise level 
on the stochastic stability of Nash equilibria. In particular, we present examples of games 
in which when the number of players or the noise level increases, a population undergoes 
a transition between its equilibria. 
\vspace{5mm}

\vspace{5mm}

\eject

\section{Introduction}

Many socio-economic and biological processes can be modeled as systems of interacting 
individuals; see for example papers in econophysics bulletin \cite{ekono}. 
One may then try to derive their global behavior from individual interactions 
between their basic entities. Such approach is fundamental in statistical physics 
which deals with systems of many interacting particles. We will explore similarities 
and differences between systems of many interacting players maximizing their individual payoffs 
and particles minimizing their interaction energy.  

Here we will consider game-theoretic models of many interacting agents \cite{wei,hof,ams}. 
In such models, agents have at their disposal certain strategies and their payoffs 
in a game depend on strategies chosen both by them and by their opponents.
A configuration of a system, that is an assignment of strategies to agents, 
is a Nash equilibrium if for any agent, for fixed strategies of his opponents, 
changing the current strategy will not increase his payoff. 
One of the fundamental problems in game theory is the equilibrium selection
in games with multiple Nash equilibria. In two-player games with two strategies 
we may have two Nash equilibria: a payoff dominant (also called efficient) 
and a risk-dominant one. In the efficient equilibrium, players 
receive highest possible payoffs. The strategy is risk-dominant 
if it has a higher expected payoff against a player playing both strategies 
with equal probabilities. It is played by individuals averse to risks. 

One of the selection methods is to construct a dynamical system 
where in the long run only one equilibrium is played with a high frequency. 
Here we will discuss adaptive dynamics where agents adapt in some optimal way 
to the environment created by other players and with a small probability, 
representing the noise of the system, they make mistakes. 
To describe the long-run behavior of such stochastic dynamics, Foster and Young \cite{foya} 
introduced a concept of stochastic stability. A configuration of a system 
is {\bf stochastically stable} if it has a positive probability in the stationary state 
of the above dynamics in the limit of zero noise. It means that in the long run 
we observe it with a positive frequency.

The main goal of this paper is to review results concerning the dependence 
of the long-run behaviour of the system on the number of players and the noise level.
We will show that in many games, when the number of players or the noise level 
increases, the population undergoes a transition between its equilibria.  

In spatial games, players are located on vertices of certain graphs and they interact only 
with their neighbors  \cite{blume1,ellis1,young2,ellis2,nowak1,nowak2,linnor,doebeli,
sabo,hauert}. In discrete moments of time, players adapt to their neighbors by choosing with 
a high probability the strategy which is the best response, i.e. the one which maximizes 
the sum of the payoffs obtained in individual games and with a small probability they make mistakes. 
Now, for any arbitrarily low but fixed noise, if the number of players is big enough, 
then the probability of any individual configuration 
is practically zero. It means that for a large number of players, to observe a stochastically 
stable configuration we must assume that players make mistakes with extremely small probabilities. 
On the other hand, it may happen that in the long run, for a low but fixed noise 
and sufficiently big number of players, the stationary state is highly concentrated 
on an ensemble consisting of one Nash configuration and its small perturbations, 
i.e. configurations where most players play the same strategy. 
We will call such configurations {\bf ensemble stable.} 
  
We will consider here games with symmetric Nash equilibria 
and homogeneous Nash configurations. By the stochastic stability of a strategy 
or a Nash equilibrium we mean the stochastic stability of the corresponding Nash 
configuration. We will present examples of spatial games with three strategies 
where concepts of stochastic stability and ensemble stability do not coincide \cite{statmech}. 
In particular, we may have the situation where a stochastically stable strategy is played 
in the long run with an arbitrarily low frequency. We discuss also an effect 
of adding a dominated strategy to a game with two strategies. In particular, 
the presence of such a strategy may cause a risk and payoff-dominant 
strategy to be observed in the long run with a frequency close to zero. 
In above models, when the number of players or the noise level increases, 
a population undergoes a transition between its equilibria.

We will also review two models of adaptive dynamics of a darwinian type where
states of a population are characterized by a number of individuals playing the first strategy. 
In both models, the selection part of the dynamics ensures that if the mean payoff 
of a given strategy at the time $t$ is bigger than the mean payoff of the other one, 
then the number of individuals playing the given strategy should increase in $t+1$. 
In the first model, introduced by Kandori, Mailath and Rob \cite{kmr}, 
one assumes (as in the standard replicator dynamics) that individuals receive average payoffs 
with respect to all possible opponents - they play against the average strategy. 
In the second model, introduced by Robson and Vega-Redondo \cite{rr,vr}, at any moment of time, 
individuals play only one game with randomly chosen opponents. In both models, 
players may mutate with a small probability hence the population may move against 
a selection pressure.

It was shown that in the Kandori-Mailath-Rob model, the risk-dominant strategy 
is stochastically stable - if the mutation level is small enough we observe it 
in the long run with the frequency close to one \cite{kmr}. 
In the model of Robson and Vega-Redondo, the payoff-dominant strategy is stochastically stable. 
It is one of very few models in which a payoff-dominant strategy is stochastically stable 
in the presence of a risk-dominant one. We showed in \cite{population} that in the sequential dynamics 
in the model of Robson and Vega-Redondo, for any arbitrarily low but a fixed level of mutations, 
if the number of players is sufficiently big, a risk-dominant strategy is played 
in the long run with a frequency close to one. More precisely, we proved that an ensemble
consisiting of states of the population in which only its small fraction play the efficient strategy
has the probability close to one in the stationary state. Again, in the limit of the infinite number 
of players, one has to consider the ensemble stability rather than the stability 
of individual configurations. 

We showed that stochastically stable efficient strategy is observed with a very low frequency. 
It means that when the number of players increases, the population undergoes a transition 
from the efficient equilibrium to the risk-dominant one.

In Section 2, we will introduce basic notions of game theory. In Section 3, 
spatial games are discussed. In Section 4, we compare ground-state configurations 
in lattice-gas models and Nash configurations in spatial games. 
In Section 5, we introduce a concept of the ensemble stability and present an example 
in which a stochastically stable strategy is played in the long run with a frequency close to zero. 
In Section 6, we review results concerning stochastic and ensemble stability in adaptive dynamics. 

\newtheorem{theo}{Theorem}
\newtheorem{defi}{Definition}
\newtheorem{hypo}{Hypothesis}
\newtheorem{prop}{Proposition}

\section{Nash equilibria}
To characterize a game-theoretic model one has to specify the set of players, 
strategies they have at their disposal and payoffs they receive.  
Although in many models the number of players is very large, 
their strategic interactions are usually decomposed into a sum of two-player games. 
Only recently there have appeared some systematic studies of truly multi-player games 
\cite{kim,broom,koba,multi,tadek}. Here we will consider only two-player games 
with two or three strategies. The payoff of any player depends not only 
on his strategy but also on strategies played by his opponents.
It can be represented by a $kxk$ matrix, where $k$ is the number of strategies.
Let us begin by describing a game with two strategies and two symmetric Nash equilibria.
A generic payoff matrix is given by
\eject

\noindent {\bf Example 1}
\vspace{3mm}

\hspace{23mm} A  \hspace{2mm} B   

\hspace{15mm} A \hspace{3mm} a  \hspace{3mm} b 

U = \hspace{6mm} 

\hspace{15mm} B \hspace{3mm} c  \hspace{3mm} d,

where the $ij$ entry, $i,j = A, B$, is the payoff of the first (row) player when
he plays the strategy $i$ and the second (column) player plays the strategy $j$. 
We assume that both players are the same and hence payoffs of the column player are given 
by the matrix transposed to $U$; such games are called symmetric. 

An assignment of strategies to both players is a Nash equilibrium, if for each player, 
for a fixed strategy of his opponent, changing the current strategy will not increase his payoff. 
If $a>c$ and $d>b$, then $(A,A)$ and $(B,B)$ are two Nash equilibria. 
If $a+b<c+d$, then the strategy $B$ has a higher expected payoff against a player playing 
both strategies with equal probabilities. We say that $B$ risk dominates the strategy $A$ 
(the notion of the risk-dominance was introduced and thoroughly studied 
by Hars\'{a}nyi and Selten \cite{hs}). If at the same time $a>d$, 
then we have a selection problem of choosing between the payoff-dominant 
(also caled efficient) equilibrium $(A,A)$ and the risk-dominant $(B,B)$.  

Games with symmetric payoff matrices are called doubly symmetric or potential games \cite{mon}. 
More generally, a game is called a {\bf potential game} if its payoff matrix can be changed 
to a symmetric one by adding payoffs to its columns. Such payoff transformation does not change 
strategic character of the game, in particular it does not change the set of its equilibria. 
More formally, it means that there exists a symmetric matrix $V$ called a potential of the game 
such that for any three strategies $A, B$, and $C,$
\begin{equation}
U(A,C)-U(B,C)=V(A,C)-V(B,C).
\end{equation} 

It is easy to see that every game with two strategies has a potential $V$ with 
$V(A,A)=a-c$, $V(B,B)=d-b$, and $V(A,B)=V(B,A)=0.$ It follows that in two-player games 
with two strategies an equilibrium is risk-dominant if and only if it has a bigger potential.
 
\section{Spatial games with local interactions}
Let $\Lambda$ be a finite subset of the simple lattice ${\bf Z}^{2}$ 
(we may think about a square centered at the origin of the lattice). 
Every site of $\Lambda$ is occupied by one player who
has at his disposal one of $k$ different pure strategies. 
Let $S$ be a set of pure strategies, then $\Omega_{\Lambda}=S^{\Lambda}$ is the set
of all possible configurations of players, that is all possible assignments 
of pure strategies to individual players. For every $i \in \Lambda$, 
$X_{i}$ is the strategy of the $i-$th player in the configuration 
$X \in \Omega_{\Lambda}$ and $X_{-i}$
denotes strategies of all remaining players; $X$ therefore can be represented 
as the pair $(X_{i},X_{-i})$. $U: S \times S \rightarrow R$ is a matrix 
of payoffs of our stage game. $U(A,B), A, B \in S$ is the payoff 
of the first (row) player playing the strategy $A$ when the second one (a column player)
is playing $B$. We will consider here only symmetric games so the payoff of
the second player is given by $U(B,A)$ (the payoff matrix of the second player 
is the transpose of the payoff matrix $U$ of the first one). 
Every player interacts only with his neighbors and his payoff 
is the sum of the payoffs resulting from individual games.
We assume that he has to use the same strategy for all neighbors. 
Let $N_{i}$ denote the neighborhood of the $i-$th player. 
For the nearest-neighbor interaction we have $N_{i}=\{j; |j-i|=1\}$,
where $|i-j|$ is the distance between $i$ and $j$.
For $X \in \Omega_{\Lambda}$ we denote by $\nu_{i}(X)$ the payoff 
of the $i-$th player in the configuration $X$:
\begin{equation}
\nu_{i}(X)=\sum_{j \in N_{i}}U(X_{i}, X_{j})
\end{equation}

\begin{defi}
$X \in \Omega_{\Lambda}$ is a {\bf Nash configuration} if for every $i \in \Lambda$
and $A \in S$ $\nu_{i}(X_{i},X_{-i}) \geq \nu_{i}(A,X_{-i})$.
\end{defi}

In this paper we will discuss only coordination games,
where there are $k$ pure symmetric Nash equilibria and therefore $k$ homogeneous
Nash configurations, where all players play the same strategy. 

We describe now the deterministic dynamics of the {\bf best-response rule}. 
Namely, at each discrete moment of time $t=1,2,...$, a randomly chosen player may update 
his strategy. He simply adopts the strategy, $X_{i}^{t}$, which gives him 
the maximal total payoff $\nu_{i}(X_{i}^{t}, X^{t-1}_{-i})$ 
for given $X^{t-1}_{-i}$, a configuration of strategies 
of remaining players at time $t-1$. 

Now we allow players to make mistakes with a small probability, 
that is to say they may not choose the best response. A probability of making
a mistake may depend on the state of the system (a configuration of strategies
of neighboring players). We will assume that this probability is a decreasing function
of the payoff lost as a result of a mistake \cite{blume1}. 
In the {\bf log-linear rule}, the probability of chosing by the $i-$th player
the strategy $X_{i}^{t}$ at time $t$ is given by the following conditional probability:

\begin{equation}
p_{i}^{T}(X_{i}^{t}|X_{-i}^{t-1})=
\frac{e^{(1/T)\nu_{i}( X_{i}^{t},X_{-i}^{t-1})}}{\sum_{A \in S}
e^{(1/T)\nu_{i}(A,X_{-i}^{t-1})}},
\end{equation}
where $T>0$ measures the noise level.

Let us observe that if $T \rightarrow 0$, 
$p_{i}^{T}$ converges to the best-response rule.
Our stochastic dynamics is an example of an ergodic Markov chain 
with $|S^{\Lambda}|$ states. Therefore, it has a unique stationary 
state which we denote by $\mu_{\Lambda}^{T}.$  

The following definition was first introduced by Foster and Young \cite{foya}:

\begin{defi}
$X \in \Omega_{\Lambda}$ is {\bf stochastically stable} 
if  $\lim_{T \rightarrow 0}\mu_{\Lambda}^{T}(X) >0.$
\end{defi} 
If $X$ is stochastically stable, then the frequency of visiting $X$ converges to
a positive number along any time trajectory almost surely. It means that in 
the long run we observe $X$ with a positive frequency. In most models it is
usually equal to $1$. 

In examples below, we consider symmetric games with symmetric Nash equilibria 
and homogeneous Nash configurations. By the stochastic stability of a strategy 
or a Nash equilibrium we mean the stochastic stability of the corresponding Nash 
configuration.

\section{Ground states and Nash configurations}

We will present here one of the basic models of interacting particles.
In classical lattice-gas models, particles occupy lattice sites and interact
only with their neighbors. The fundamental concept
is that of a ground-state configuration. It can be formulated conveniently
in the limit of an infinite lattice (the infinite number of particles).
Let us assume that every site of the ${\bf Z}^{2}$ lattice can be occupied 
by one of $k$ different particles. An infinite-lattice configuration 
is an assignment of particles to lattice sites, i.e., an element of $\Omega =
\{1,...,k\}^{{\bf Z}^{2}}$. If $X \in \Omega$ and $i \in  {\bf
Z}^{2}$, then we denote by $X_{i}$ a restriction of $X$ to $i$.
We will assume here that only nearest-neighbor particles interact.
The energy of their interaction is given by a symmetric $k \times k$ matrix $V$.
An element $V(A,B)$ is the interaction energy of two nearest-neighbor
particles of type $A$ and $B$. The total energy of a system 
in a configuration $X$ in a finite region $\Lambda$ can be then written as 
$H_{\Lambda}(X)=\sum_{(i,j) \in \Lambda} V(X_{i},X_{j})$.

$Y$ is a {\bf local excitation} of $X$, $Y \sim X$, $Y,X \in
\Omega$ , if there exists a finite $\Lambda \subset {\bf Z}^{d}$
such that $X = Y$ outside $\Lambda.$ 

For $Y \sim X$, the {\bf relative energy} is defined by 
\begin{equation}
H(Y,X)=\sum_{(i,j)} (V(Y_{i},Y_{j})-V(X_{i},X_{j})),
\end{equation}
where the summation is with respect to pairs of nearest neighbors
on ${\bf Z}^{2}$. Observe that this is the finite sum; the energy difference
between $Y$ and $X$ is equal to 0 outside some finite $\Lambda$.

\begin{defi}
$X \in \Omega$ is a {\bf ground-state configuration} of $V$ if 
$$H(Y,X) \geq 0 \; \; for \; \; any \; \; Y \sim X.$$  
\end{defi}
That is, we cannot lower the energy of a ground-state
configuration by changing it locally.

The energy density $e(X)$ of a configuration $X$ is 
\begin{equation}
e(X)=\liminf_{\Lambda \rightarrow {\bf Z}^{2}}
\frac{H_{\Lambda}(X)}{|\Lambda|},
\end{equation}
where $|\Lambda|$ is the number of lattice sites in $\Lambda$. It
can be shown that any  ground-state configuration has the minimal
energy density \cite{sinai}. It means that local conditions present in the definition 
of a ground-state configuration force global minimization of the energy density.

As we see, the concept of a ground-state configuration is very similar
to that of a Nash configuration. We have to identify
particles with agents, types of particles with strategies
and instead of minimizing interaction energies we should maximize payoffs.
There are however profound differences. First of all, 
ground-state configurations can be defined only for symmetric matrices; 
an interaction energy is assigned to a pair of particles, payoffs are assigned 
to individual players and may be different for each of them.
Ground-state configurations are stable with respect to all local changes, 
Nash configurations are stable only with respect to one-player changes.
It means that for the same symmetric matrix $U$, there may exist a configuration 
which is a Nash configuration but not a ground-state configuration 
for the interaction marix $-U$. The simplest example is given by the following matrix:

\vspace{5mm}

\noindent {\bf Example 2}

\hspace{23mm} A \hspace{2mm} B   

\hspace{15mm} A \hspace{3mm} 2 \hspace{3mm} 0 

U = \hspace{6mm} 

\hspace{15mm} B \hspace{3mm} 0 \hspace{3mm} 1 

\noindent $(A,A)$ and $(B,B)$ are Nash configurations for a system consisting 
of two players but only $(A,A)$ is a ground-state configuration for $V=-U.$  
We may therefore consider the concept of a ground-state configuration 
as a refinement of a Nash equilibrium.

For any classical lattice-gas model there exists at least one 
ground-state configuration. This can be seen in the following way.
We start with an arbitrary configuration. If it cannot be changed locally
to decrease its energy it is already a ground-state configuration.
Otherwise we may change it locally and decrease the energy of the system.
If our system is finite, then after a finite number of steps we arrive at a
ground-state configuration; at every step we decrease the energy of the system
and for every finite system its possible energies form a finite set.
For an infinite system, we have to proceed ad infinitum converging
to a ground-state configuration (this follows from the compactness of $\Omega$).
Game models are different. It may happen that a game with a nonsymmetric 
payoff matrix may not posess a Nash configuration. The classical example is that 
of the Rock-Scissors-Paper game given by the following matrix.

\vspace{5mm}

\noindent {\bf Example 3}

\hspace{23mm} R  \hspace{2mm} S \hspace{2mm} P  

\hspace{15mm} R  \hspace{3mm} 1  \hspace{3mm} 2 \hspace{3mm} 0

U = \hspace{6mm} S \hspace{3mm} 0  \hspace{3mm} 1 \hspace{3mm} 2

\hspace{15mm} P \hspace{3mm} 2  \hspace{3mm} 0 \hspace{3mm} 1

One may show that this game dos not have any Nash configurations on ${\bf Z}$
and ${\bf Z}^{2}$ but many Nash configurations on the triangular lattice.

In short, ground-state configurations minimize the total energy of a particle system, 
Nash configurations do not necessarily maximize the total payoff of a society.

Ground-state configuration is an equilibrium concept for
systems of interacting particles at zero temperature. For positive temperatures,
we must take into account fluctuations caused by thermal motions of particles.
Equilibrium behavior of the system results then from the competition between
its energy $V$ and entropy $S$ (which measures the number 
of configurations corresponding to a macroscopic state), i.e. the minimization 
of its free energy $F=V-TS$, where $T$ is the temperature 
of the system - a measure of thermal motions. At the zero temperature, 
$T=0$, the minimization of the free energy reduces to the minimization of the energy.
This zero-temperature limit looks very similar to the zero-noise limit present 
in the definition of the stochastic stability. Equilibrium behavior of a system 
of interacting particles can be described by specifying probabilities 
of occurence for all particle configurations. More formally, it is described 
by a Gibbs state (see \cite{geo} and references therein).

We construct it in the following way. Let
$\Lambda$ be a finite subset of ${\bf Z}^{2}$ and  $\rho^{T}_{\Lambda}$ 
the following probability mass function on
$\Omega_{\Lambda}=(1,...,k)^{\Lambda}$: 
\begin{equation}
\rho_{\Lambda}^{T}(X)=(1/Z^{T}_{\Lambda})\exp(-H_{\Lambda}(X)/T),
\end{equation}
for every $X \in \Omega_{\Lambda}$, where
\begin{equation}
Z^{T}_{\Lambda}=\sum_{X \in \Omega_{\Lambda}}\exp(-H_{\Lambda}(X)/T)
\end{equation}
is a normalizing factor and $T$ is the temperature of the system.

We define a {\bf Gibbs state} as a limit of $\rho^{T}_{\Lambda}$ as
$\Lambda \rightarrow {\bf Z}^{2}$. One can prove
that a limit of a translation-invariant Gibbs state for a given interaction
as $T \rightarrow 0$ is a measure supported by ground-state configurations.
One of the fundamental problems of statistical mechanics is a characterization
of low-temperature Gibbs states for given interactions between particles.

Let us now come back to spatial games with players located on a finite region
$\Lambda$ of the ${\bf Z^{2}}$ lattice and receiving a payoff given by the matrix in Example 1. 
It has two homogeneous  Nash configurations, $X^{A}$ and $X^{B}$, 
in which all players play the same strategy $A$ or $B$ respectively. 
If $V$ is a potential of the stage game, then $\sum_{(i,j)\in \Lambda}V(X_{i},X_{j})$ 
is a potential of a configuration $X$ in the corresponding spatial game.
One can show \cite{young2} that  

\begin{equation}
\mu^{T}_{\Lambda}(X)=\frac{e^{(1/T)\sum_{(i,j)\in \Lambda}V(X_{i},X_{j})}}
{\sum_{Z \in \Omega_{\Lambda}}e^{(1/T)\sum_{(i,j)\in \Lambda}V(Z_{i},Z_{j})}}.
\end{equation}
is the unique stationary state of the log-linear dynamics.

We may now explicitly perform the limit $T \rightarrow 0$
and use (1) to obtain that the risk-dominant configuration $X^{B}$
is stochastically stable. 

\section{Ensemble stability in spatial games}

The concept of stochastic stability involves individual configurations of players. 
In the zero-noise limit, the stationary state is usually concentrated on one 
or at most few configurations. However, for a low but fixed noise and
for a big number of players, the probability of any individual configuration 
of players is practically zero. The stationary state, however, may be highly 
concentrated on an ensemble consisting of one Nash configuration and its small 
perturbations, i.e., configurations, where most players play the same strategy. 
Such configurations have relatively high probability in the stationary state. 
We call such configurations {\bf ensemble stable} \cite{statmech}.

\begin{defi}
$X \in \Omega_{\Lambda}$ is {\bf $\epsilon$-ensemble stable}
if $\mu_{\Lambda}^{T}(Y \in \Omega_{\Lambda};Y_{i} \neq X_{i}) < \epsilon$
for any $i \in \Lambda$ if $\Lambda \supset \Lambda(T)$ for some $\Lambda(T)$. 
\end{defi}

\begin{defi}
$X \in \Omega_{\Lambda}$ is {\bf low-noise ensemble stable}
if for every $\epsilon>0$ there exists $T(\epsilon)$ such that if
$T<T(\epsilon)$, then $X$ is $\epsilon$-ensemble stable.
\end{defi}

If $X$ is $\epsilon$-ensemble stable for some $T>0$ and $\epsilon$ close to zero, then the ensemble 
consisting of $X$ and configurations which are different from $X$ at at most few sites has 
the probability close to one in the stationary state. It does not follow,
however, that $X$ is necessarily low-noise ensemble or stochastically stable as it happens 
in the following example.

Players are located on a finite subset $\Lambda$ of ${\bf Z}^{2}$ 
(with periodic boundary conditions) and interact with their four nearest neighbors. 
They have at their disposal three strategies: 
$A, B,$ and $C$. The payoffs are given by the following symmetric matrix:
\vspace{2mm}

\noindent {\bf Example 4}
\vspace{3mm}

\hspace{23mm} A  \hspace{5mm} B \hspace{5mm} C  

\hspace{15mm} A \hspace {3mm} 0  \hspace{5mm} 0.1 \hspace{4mm} 1

U = \hspace{6mm} B \hspace{3mm} 0.1  \hspace{1mm} $2+\alpha$ \hspace{1mm} 1

\hspace{15mm} C \hspace{3mm} 1  \hspace{5mm} 1 \hspace{6mm} 2,

where $\alpha \geq 0$.
\vspace{3mm}

Our game has two Nash equilibria: $(B,B)$ and $(C,C)$, and the corresponding spatial game has 
two homogeneous Nash configurations: $X^{B}$ and $X^{C}.$ For $\alpha=0$, 
both $X^{B}$ and $X^{C}$ are ground state-configurations for $-U$; 
for $\alpha>0$ only $X^{B}$ is a ground-state configuration. 

Observe that the strategy $A$ gives a player the lowest payoff regardless 
of a strategy chosen by his opponent. Such strategy is called {\bf dominated}.
It is easy to see that dominated strategies cannot be present in any Nash equilibrium. 
Therefore such strategies should not be used by players and consequently we might think 
that their presence should not have any impact on the long-run behavior of the system. 
We will see that this is not true in Example 4.

The unique stationary state of the log-linear dynamics (3) is  given in (8) with $V=U$.
Let us start our discussion with $\alpha=0.$
It follows from (8) that $\lim_{T \rightarrow 0}\mu_{\Lambda}^{T}(X^{A^{i}})=1/2$, 
$A^{i}=B, C$ so $B$ and $C$ are stochastically stable.
Let us investigate the long-run behavior of our system for large $\Lambda$, 
that is for a big number of players. Observe that 
$\lim_{\Lambda \rightarrow {\bf Z}^{2}}\mu_{\Lambda}^{T}(X)=0$ 
for every $X \in \Omega = S^{{\bf Z}^{2}}$.
Hence for large $\Lambda$ and $T>0$ we may only observe,
with reasonable positive frequencies, ensembles of configurations 
and not particular configurations. We will be interested in ensembles 
which consist of a Nash configuration and its small perturbations, 
that is configurations, where most players adopt the same strategy. 
We perform first the limit $\Lambda \rightarrow {\bf Z}^{2}$
and obtain an infinite volume Gibbs state

\begin{equation}
\mu^{T} = \lim_{\Lambda \rightarrow {\bf Z}^{2}}\mu_{\Lambda}^{T}
\end{equation}
We may then apply a technique developed by Bricmont and Slawny \cite{brsl1,brsl2}.
They studied low-temperature stability of the so-called dominant 
ground-state configurations. It follows from their results that

\begin{equation}
\mu^{T}(X_{i}=C)>1-\epsilon(T)  
\end{equation}
for any $i \in {\bf Z}^{2}$ and $\epsilon(T) \rightarrow 0$ as $T \rightarrow 0$ \cite{statmech}. 

The following theorem is a simple consequence of (10).
\begin{theo}
If $\alpha =0$, then $X^{C}$ is low-noise ensemble stable.
\end{theo} 

We see that for any low but fixed $T$, if the number of players is big enough,
then in the long run, almost all players use $C$ strategy. 
On the other hand, if for any fixed number of players, $T$ is lowered substantially,
then $B$ and $C$ appear with frequencies close to $1/2$.

Let us sketch briefly the reason of such a behavior.
While it is true that both $X^{B}$ and $X^{C}$ have the same potential 
which is the half of the payoff of the whole system (it plays the role 
of the total energy of a system of interacting particles), 
the $X^{C}$ Nash configuration has more lowest-cost excitations. 
Namely, one player can change its strategy and switch to either
$A$ or $B$ and the total payoff will decrease by $8$ units. Players 
in the $X^{B}$ Nash configuration have only one possibility, 
that is to switch to $C$; switching to $A$ decreases the total payoff by $15.2$. 
Now, the probability of the occurrence of any configuration in the Gibbs state 
(which is the stationary state of our stochastic dynamics) 
depends on the total payoff in an exponential way. 
One can prove that the probability of the ensemble consisting of the $X^{C}$ Nash 
configuration and configurations which are different from it 
at few sites only is much bigger than the probability of the analogous
$X^{B}$-ensemble. It follows from the fact that the $X^{C}$-ensemble 
has many more configurations than the $X^{B}$-ensemble. On the other hand,
configurations which are outside $X^{B}$ and $X^{C}$-ensembles 
appear with exponentially small probabilities. It means that for large enough systems 
(and small but not extremely small $T$) we observe in the stationary state the $X^{C}$ 
Nash configuration with perhaps few different strategies. The above argument was made 
into a rigorous proof for an infinite system of the closely related lattice-gas model 
(the Blume-Capel model) of interacting particles by Bricmont and Slawny in \cite{brsl1}.  

For $\alpha=0$, both $X^{B}$ and $X^{C}$ have the same total payoff. 
$X^{C}$ has lowest-cost fluctuations and therefore it is low-noise ensemble stable. 
For $\alpha>0$,  $X^{C}$ has a smaller total payoff but nevertheless 
one can prove \cite{statmech} that in the long run $C$ is played with a frequency close to $1$ 
if the noise level is low but not extremely low. 

\begin{theo}
For every $\epsilon>0$, there exist $\alpha(\epsilon)$ and $T(\epsilon)$ 
such that for every $0<\alpha<\alpha(\epsilon)$, 
there exists $T(\alpha)$ such that for $T(\alpha)<T<T(\epsilon)$, 
$X^{C}$ is $\epsilon$-ensemble stable,
and for $0<T<T(\alpha)$, $X^{B}$ is $\epsilon$-ensemble stable.
\end{theo}

Observe that for $\alpha=0$, both $X^{B}$ and $X^{C}$ are stochastically stable
(they appear with the frequency $1/2$ in the limit of zero noise) but $X^{C}$
is low-noise ensemble stable. For small $\alpha > 0$, $X^{B}$ is both stochastically
(it appears with the frequency $1$ in the limit of zero noise) and low-noise 
ensemble stable. However, for intermediate noise 
$T(\alpha)<T<T(\epsilon)$, if the number of players is big enough, then in the long run, 
almost all players use $C$ strategy - $X^{C}$ is ensemble stable). 
If we lower T below $T(\alpha)$, then almost all players start to use $B$ strategy. 
We may say that at $T=T(\alpha)$ the society of players undergoes a {\bf phase transition} 
from $C$ to $B$-behavior. 

Stochastic and ensemble stability of three-player spatial games were discussed in \cite{physica} 
and of some other spatial games in \cite{statphys}.

\section{Stochastic and ensemble stability in adaptive dynamics}

Here we will review two models of darwinian adaptive dynamics. 
In both of them, the selection part of the dynamics ensures that if 
the mean payoff of a given strategy at the time $t$ is bigger than the mean payoff 
of the other one, then the number of individuals playing the given strategy should increase 
in $t+1$. In the first model, introduced by Kandori, Mailath, and Rob \cite{kmr}, one assumes 
(as in the standard replicator dynamics \cite{wei,hof}) that individuals receive 
average payoffs with respect to all possible opponents - they play against the average strategy. 
In the second model, introduced by Robson and Vega-Redondo \cite{rr,vr}, at any moment of time, 
individuals play only one game with randomly chosen opponents. In both models, players may mutate 
with a small probability hence the population may move against a selection pressure. 

In the above models, at every discrete moment of time $t$, the state of the population
is described by the number of individuals, $z_{t}$, playing the strategy $A$ in the two-player 
symmetric game with the payoff matrix given in Example 1. Formally, by the state space 
we mean the set $\Omega=\{z, 0 \leq z\leq n\}.$ Now we will describe the dynamics of our system. 
It consists of two components: selection and mutation. Let $\pi_{i}(z_{t}), i=A,B$, 
denote the mean payoff of a strategy at the time $t$. 
In their paper \cite{kmr}, Kandori, Mailath, and Rob  write
\begin{equation}
\pi_{A}(z_{t})=\frac{a(z_{t}-1)+b(n-z_{t})}{n-1},
\end{equation} 
$$\pi_{B}(z_{t})=\frac{cz_{t}+d(n-z_{t}-1)}{n-1},$$
provided $0<z_{t}<n$.

It means that in every time step, players are paired infnitely many times
to play the game or equivalently, each player plays with every other player 
and his payoff is the sum of corresponding payoffs. 

The selection dynamics is formalized in the following way:

\begin{equation}
z_{t+1} > z_{t} \hspace{2mm} if \hspace{2mm} \pi_{A}(z_{t}) > \pi_{B}(z_{t}),
\end{equation}
$$z_{t+1} < z_{t} \hspace{2mm} if \hspace{2mm} \pi_{A}(z_{t}) < \pi_{B}(z_{t}),$$
$$z_{t+1}= z_{t} \hspace{2mm} if \hspace{2mm} \pi_{A}(z_{t}) = \pi_{B}(z_{t}),$$
$$z_{t+1}= z_{t} \hspace{2mm} if \hspace{2mm} z_{t}=0  \hspace{2mm} or \hspace{2mm} z_{t}=n.$$

Now mutations are added. At every moment $t$, each player switches 
to a new strategy with some probability $\epsilon$. It is easy to see 
that for any two states of the population, there is a positive probability
of the transition between them in some finite number of time steps. We have therefore obtained
an irreducible Markov chain with $n+1$ states. It has a unique stationary state 
(a probability mass function) which we denote by $\mu^{\epsilon}_{n}.$
The following theorem was proved in \cite{kmr}. 

\begin{theo}
For any large enough $n$, $\lim_{\epsilon \rightarrow 0}\mu^{\epsilon}_{n}(0)=1$ 
in the average strategy (K-M-R) model.
\end{theo}

It means that in the long run, in the limit of no mutations, all players play 
the risk-dominant strategy $B$, that is $B$ is stochastically stable. 

We will outline the proof of the above theorem (see \cite{kmr,vr}). 
It is based on the tree representation of stationary states of irreducible Markov chains
\cite{freiwen1,freiwen2,shub} (see also the Appendix). 

We will first show that $z=0$ and $z=n$ and possibly $nx^{*}$ (if it is an integer) 
are the only absorbing states of the mutation-free dynamics (the selection part). 
Let $U_{A}$ and $U_{B}$ are the basins of attraction of $z=n$ and $z=0$ respectively.
The basins do not overlap and we have that $U_{A}=\{k, nx^{*}< k \leq n\}$ and 
$U_{B}=\{k, 0 \leq k < nx^{*}\}$, where $x^{*}=(d-b)/(d-b+a-c)$ 
is the solution of the equation $\pi_{A} = \pi_{B},$ 
that is $x^{*}$ is the mixed Nash equilibrium of the game. 
If $nx^{*}$ is an integer, it is the additional invariant state of the dynamics.

Now to prove Theorem 3 it is enough to notice that $x^{*}>1/2$ for the payoff matrix in Example 1 
and therefore the population needs more mutations to move from $z=0$ to $z=n$ 
than the other way around. Therefore $q_{m}(z=0)$ has the lower order in $\epsilon$ in (16) 
than $q_{m}(z=n)$ so $B$ is stochastically stable.  
\vspace{3mm}

The general set up in the model of Robson and Vega-Redondo \cite{rr,vr} is the same.
However, individuals are paired only once at every time step and play only one game
before the selection process takes place. Let $p_{t}$ denote the random variable
which describes the number of cross-pairings, i.e. the number of pairs 
of matched individuals playing different strategies at the time $t$.
Let us notice that $p_{t}$ depends on $z_{t}$.
For a given realization of $p_{t}$ and $z_{t}$,
mean payoffs obtained by each strategy are as follows:
\begin{equation}
\tilde{\pi}_{A}(z_{t},p_{t})=\frac{a(z_{t}-p_{t})+bp_{t}}{z_{t}},
\end{equation} 
$$\tilde{\pi}_{B}(z_{t},p_{t})=\frac{cp_{t}+d(n-z_{t}-p_{t})}{n-z_{t}},$$
provided $0<z_{t}<n$. 

Let us denote by $\tilde{\mu}^{\epsilon}_{n}$
the stationary state of the corresponding Markov chain.
It was proved in \cite{rr,vr} that the payoff-dominant strategy 
is stochastically stable.

\begin{theo}
For any large enough $n$, $\lim_{\epsilon \rightarrow 0} \tilde{\mu}^{\epsilon}_{n}(n)=1$ 
in the random matching (R-V-R) model.
\end{theo}

Let us again outline the proof.
Due to the stochastic nature of matching, basins of attraction overlap here. 
First of all, one can show that there exists $k$ such that 
if $n$ is large enough and $z_{t} \geq k$, 
then there is a positive probability (a certain realization of $p_{t}$)
that after a finite number of steps of the mutation-free selection dynamics,
all players will play $A$. Likewise, if $z_{t} <k$ (for any $k \geq 1$, then if the number of players 
is large enough, then after a finite number of steps of the mutation-free selection dynamics 
all players will play $B$. In other words, $z=0$ and $z=n$ 
are the only absorbing states of the mutation-free dynamics and there are no other recurrent classes.
Moreover, if $n$ is large enough, then if $z_{t} \geq n-k$, then the mean payoff obtained
by $A$ is always (for any realization of $p_{t}$) bigger than the mean payoff obtained by $B$.
(in the worst case all $B$-players play with $A$-players).
Therefore the size of the basin of attraction of the state $z=0$ is at most $n-k-1$
an that of $z=n$ is at least $n-k$. It follows that the system needs at least $k+1$ 
mutations to evolve from $z=n$ to $z=0$ and at most $k$ mutations to evolve from $z=0$ to $z=n$. 
Now $q_{m}(z=n)$ has the lower order in $\epsilon$ in (16) than $q_{m}(z=0)$ 
which finishes the proof.
\vspace{3mm}

Stochastic stability is concerned with the limit of no mutations. 
We have shown \cite{population} that for a low but fixed level of mutations 
in the sequential dynamics of the random matching model, if the number of players is large enough, 
then in the long run, the stochastically stable 
efficient strategy is played with the frequency close to zero. 
It is again the risk-dominant strategy, as in the Kandori-Mailath-Rob model, 
which is played with the frequency close to one. 
It means that when the number of players increases, the population undergoes a transition 
between its two equilibria. Of course, if the noise level 
is decreased for a new larger population, it settles again in the efficient equilibrium - 
after all the efficient equilibrium is stochastically stable. 
We have therefore proved in \cite{population} that two limits of stationary distributions, 
the zero mutation and the infinite number of players limit are concentrated 
on two different equilibria.  

Let us describe our model in more detail. In sequential dynamics, in one time unit, only
one player can change his strategy. The number of $A$-players in the population can increase by one
in $t+1$, if a $B$-player is chosen in $t$ which happens with 
the probability $(n-z_{t})/n$. Analogously, the number of $B$-players 
in the population can increase by one in $t+1$, if an $A$-player 
is chosen in $t$ which happens with the probability $(z_{t})/n$. 

The player who has a revision opportunity will chose in $t+1$ 
with the probability $1-\epsilon$ the strategy with a higher average payoff in $t$ 
and the other one with the probability $\epsilon$. 

The following theorem was proven in \cite{population}.

\begin{theo}
In the sequential dynamics of the random matching model, for any $\delta >0$ 
and $\beta >0$ there exist $\epsilon(\delta, \beta)$
and $n(\epsilon)$ such that for any $n > n(\epsilon)$
$$\tilde{\mu}_{n}^{\epsilon}(z<\beta n) > 1- \delta.$$
\end{theo}

Let us note that the above theorem concerns an ensemble of configurations,
not an individual one. In the limit of the infinite number of players, that is 
the infinite number of configurations, every single configuration has zero probability
in the stationary state. It is an ensemble of configurations that might be stable.

We also showed in \cite{population} that in the random matching model, 
stochastic stability itself may depend on the number of players.
If the population consists of only one $B$-player and $n-1$ $A$-players 
and if $c>[a(n-2)+b]/(n-1)$, that is $n< (2a-c-b)/(a-c)$, 
then  $\tilde{\pi}_{B}> \tilde{\pi}_{A}.$ It means that one needs only one mutation 
to evolve from $z=n$ to $z=0.$ It is easy to see that two mutations are necessary
to evolve from $z=0$ to $z=n$. Using again the tree representation of stationary states,
(see the Appendix) one can prove the following theorem.

\begin{theo}
If $n<\frac{2a-c-b}{a-c}$, then $B$ is stochastically stable.
\end{theo}     

We already know that for any large enough $n$, efficient strategy $A$ is stochastically stable.
Again, the population changes its behavior when the number of players increases.
However, the nature of this transition is different from the one described before.
When the number of players increases, in order to see stochastically stable efficient strategy,
the mutation level should decrease substantially. 

\vspace{3mm}

\noindent {\bf Acknowledgments}: I thank Christian Maes and Joseph Slawny
for useful conversations and Rudolf Krejcar for writing Monte-Carlo programs
to simulate stochastic dynamics of spatial games. I would like to thank J\"{o}rgen Weibull 
and Peyton Young for illuminating discussions on evolutionary game theory.
Financial support by the Polish Committee for Scientific Research
under the grant KBN 5 P03A 025 20 is kindly acknowledged.

\section{Appendix}
The following tree representation of stationary distributions 
of Markov chains was proposed by Freidlin and Wentzell in \cite{freiwen1,freiwen2} 
(see also \cite{shub}). Let $(\Omega,P)$ be an irreducible Markov chain with a state space 
$\Omega$ and transition probabilities given by $P: \Omega \times \Omega \rightarrow [0,1]$. 
It has a unique stationary distribution $\mu$ (called also a stationary state). 
For $X \in \Omega$, let X-tree be a directed graph 
on $\Omega$ such that from every $Y \neq X$ there is a unique path to $X$
and there are no outcoming edges out of $X$. Denote by $T(X)$ the set of all X-trees
and let 
\begin{equation}
q(X)=\sum_{d \in T(X)} \prod_{(Y,Y') \in d}P(Y,Y'),
\end{equation}
where the product is with respect to all edges of $d$. 
Now one can show \cite{freiwen2} that
\begin{equation}
\mu(X)=\frac{q(X)}{\sum_{Y \in \Omega}q(Y)}
\end{equation}
for all $X \in \Omega.$

A state is an absorbing one if it attracts nearby states in the noise-free dynamics. 
We asume that after a finite number of steps of the noise-free dynamics we arrive 
at one of the absorbing states (there are no other recurrence classes) and stay there forever. 
Then it follows from the above tree representation that any state different from absorbing states 
has zero probability in the stationary distribution in the zero-noise limit. 
Moreover, in order to study the zero-noise limit of the stationary distribution, 
it is enough to consider paths between absorbing states. More precisely, 
we construct X-trees with absorbing states  as vertices; the family of such 
$X$-trees is denoted by $\tilde{T}(X)$. Let 
\begin{equation}
q_{m}(X)=max_{d \in \tilde{T}(X)} \prod_{(Y,Y') \in d}\tilde{P}(Y,Y'),
\end{equation}
where $\tilde{P}(Y,Y')= max \prod_{(W,W')}P(W,W')$,
where the product is taken along any path joining $Y$ with $Y'$ and the maximum 
is taken with respect to all such paths. 
Now we may observe that if $lim_{\epsilon \rightarrow 0} q_{m}(Y)/q_{m}(X)=0,$
for any $Y\neq X$, then $X$ is stochastically stable. Therefore we have to compare  
trees with the biggest products in (16); such trees we call maximal.

\end{document}